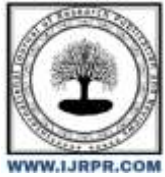

# International Journal of Research Publication and Reviews

# Evaluating the Performance of Nigerian Lecturers using Multilayer Perceptron

*I. E. Ezeibe, S.O. Okide, D.C. Asogwa*

*fnnobiefuna@gmail.com*, Department of Computer Science, Nnamdi Azikiwe University, Nigeria
*so.okide@unizik.edu.ng*, Department of Computer Science, Nnamdi Azikiwe University, Nigeria
*dc.asogwa@unizik.edu.ng*, Department of Computer Science, Nnamdi Azikiwe University, Nigeria

## ABSTRACT

Evaluating the performance of a lecturer has been essential for enhancing teaching quality, improving student learning outcomes, and strengthening the institution's reputation. The absence of such system brings about lecturer performance evaluation which was neither comprehensive nor holistic. This system was designed using a web-based platform, created a secure database, and by using a custom dataset, captured some performance metrics which included student evaluation scores, Research Publications, Years of Experience, and Administrative Duties. Multilayer Perceptron (MLP) algorithm was utilized due to its ability to process complex data patterns and generates accurate predictions in a lecturer's performance based on historical data. This research focused on designing multiple performance metrics beyond the standard ones, incorporating student participation, and integrating analytical tools to deliver a comprehensive and holistic evaluation of lecturers' performance and was developed using Object-Oriented Analysis and Design (OOAD) methodology. Lecturers' performance is evaluated by the model, and the evaluation accuracy is about 91% compared with actual performance. Finally, by evaluating the performance of the MLP model, it is concluded that MLP enhanced lecturer performance evaluation by providing accurate predictions, reducing bias, and supporting data-driven decisions, ultimately improving the fairness and efficiency of the evaluation process. The MLP model's performance was evaluated using Mean Squared Error (MSE) and Mean Absolute Error (MAE), achieved a test loss (MSE) of 256.99 and a MAE of 13.76, and reflected a high level of prediction accuracy. The model also demonstrated an estimated accuracy rate of approximately 96%, validated its effectiveness in predicting lecturer performance.

Keywords: Artificial Neural Network (ANN), Prediction, Machine Learning, Lecturer Performance, Multilayer Perceptron (MLP).

## Introduction

A system designed to predict the performance of lecturers over a specific period could contribute to a more comprehensive and holistic evaluation process. By analyzing various data points such as student feedback, teaching methodologies, course outcomes, and other relevant metrics, this system could provide insights into a lecturer's performance. This can help ensure that lecturers who lack in some areas are not only given measures to take to improve but also become skilled at engaging and educating students. (Igbokekwe, Ugo-Okoro, & Agbonye, 2015) agreed that currently Nigerian universities lack a well-rounded system for evaluating the performance of lecturers and the existing methods often fail to capture the full spectrum of a lecturer's responsibilities, which extend beyond simply teaching a course. According to (Eze, 2006), it is widely recognized that Nigeria faces a serious problem with corruption. This corruption has unfortunately spread to the university system, which is unsurprising given that universities reflect the broader society. (Okojie, 2012) also expressed concern that corruption in Nigeria has become alarming, and that the university system, as an essential part of society, cannot avoid this threat. The result is the gradual erosion of the values that are necessary for a civil and prosperous society. Universities have traditionally been seen as institutions of knowledge and integrity, shaping students into upstanding individuals. However, (Odunaya and Olujunwon, 2010) opined that the societal problems of greed, materialism, and taking shortcuts have infiltrated these institutions, leading to a worrying prevalence of corruption. This decline makes perfect sense as the National Policy on Education in Nigeria FRN stated that the quality of a nation's education system is directly linked to the quality of its teachers. The significant gap in the existing evaluation of a lecturer's performance highlights the strong need for developing a robust performance evaluation system for lecturers. It is challenging to comprehensively evaluate a lecturers' performance without a system that considers all aspects of their role.

## Related Works

In this study, (Sudiyono & Mulyasa, 2020) discussed the process and results of lecturer performance appraisals carried out by the Internal Quality Assurance Unit at universities in the city of Bandung, West Java Province – Indonesia. The objective was to examine and describe both the procedure and the results of lecturer performance assessments. The researchers utilized a descriptive method with a qualitative approach, involving stages such as observation, exploration, and member-check. Data collection techniques included documentation study, interviews, and observations. Meanwhile, the evaluation process of lecturers in the studied universities by the authors was systematically conducted in line with predefined dimensions and stages, involving input from lecturers and aiming for positive lecturer performance outcomes. Based on these findings, it was recommended that assessments of lecturer performance should adhere to procedural steps, maintain objectivity, and focus on achieving desired evaluation outcomes.

In this study, (Adigun, Irunokhai, Onihunwa, Sada, Jeje, Areo, & Ilori, 2022) developed and tested a web-based system Students' Appraisal on Teaching Performance (SATP) for students to anonymously evaluate their lecturers. The authors utilized a two-step approach. Firstly, they employed an experimental domain-driven methodology, likely an adaptation of Domain-Driven Design, to design the core framework for the SATP system. Secondly, the authors implemented the designed framework by developing a prototype (SATP) and used a Quasi-experimental methodology. This two pronged approach helped ensure the system's design aligns with the evaluation process and then assessed its effectiveness in a practical context. This research offers a valuable tool for universities and provides a reliable way for administrators to get feedback on lecturers directly from students, the core recipients of their teaching.

In this research, (Stirruph & Omade, 2021) presented a unique approach to lecturer evaluation in Nigerian universities. While existing systems often emphasize on lecturer self-evaluations, this study breaks new ground by developing a web-based student evaluation platform – Student Appraisal on Teaching Performance (SATP). This system offers valuable insights into lecturer performance directly from students, who are key stakeholders in the learning process. This study employed a descriptive survey design and data analysis relied primarily on descriptive statistics. Furthermore, this research highlights the limitations of current methods. Moving forward, a comprehensive approach – the 360- degree evaluation technique should be considered for adoption by the Nigerian universities. To the best of our current understanding, this study represents the initial effort made in uniquely improving lecturer evaluations by embracing a student evaluation system like Student Appraisal on Teaching Performance (SATP).

In this study, (Saaludin, Ismail, Mat & Harun, 2019) utilized the Analytic Hierarchy Process (AHP) to determine the importance of various criteria used in lecturer evaluations. AHP offers a structured approach to decision making by helping identify the best option, in this case, the overall score for a lecturers' evaluation. It also established a framework for setting priorities through pairwise comparisons. This involves comparing each criterion against every other one to determine their relative importance. A scale of 1 to 9 is used to represent the relative importance within each comparison. While pairwise comparisons seem straight forward, the level of confidence in their consistency can decrease as the number of comparisons increases. Therefore, a consistency ratio (CR) is calculated to assess the reliability of the results. The finding shows that among the evaluated criteria, teaching delivery holds the greatest significance. To ensure high-quality teaching, lecturers must possess attributes such as expertise, enthusiasm, and creativity in employing diverse teaching methodologies.

In this study, (Abdulkadir, Mukail, & Bitrus, 2022) utilized the Structural System Analysis and Design Methodology (SSADM), which offers a systematic approach to analyzing and designing information systems. The process involved several stages, including feasibility study, investigation of the current environment, and definition of requirements, system design, logical design, and physical design, in order to develop the application. The study's findings unveiled that none of the ten assessed aspects of lecturers' teaching received an excellent rating. It is anticipated that lecturers will sustain their identified strengths as illuminated by the empirical data. Moreover, the evaluation outcomes have identified specific areas of knowledge and skill enhancement for lecturer to consider moving forward.

This study done by (Bemile, Jackson, & Ofosu, 2014) followed a qualitative approach, inspired by the work of (Neuman, 2007). The researchers focused on a group within the Quality Assurance Unit of Methodist University College Ghana (MUCG). With permission, interviews were recorded on a smart phone and later transcribed. By analyzing the discussions, the researchers were able to draw conclusions about the Course and Lecturer Evaluation by Students (CLES) exercise. The researchers also realized that it is extremely difficult for only two staff of the Quality Assurance Unit to be at all the venues at the same time to collect the data.

This study conducted by (Isaeva, Kotliarenko, & Cherkasova, 2021) believed that video analysis and participant observation are two methods that can effectively identify changes in professors; competencies. These methods revealed new aspects of teaching behavior and communication with students that was not present before. This suggested a shift in the overall system of professional competencies for professors due to digital transformation. The researchers found out that leveraging modern digital technology, they could virtually attend about ten classes within two hours. The researchers further proposed that implementing regular "Contests of Pedagogical Excellence" could significantly improve professor evaluation procedures. This approach has the potential to benefit a wider range of educators.

This research done by (Rakhmadani & Adhinata, 2021) utilized a numerical rating scale within a web-based lecturer performance appraisal system implemented by Telkom Institute of Technology Purwokerto. This numerical rating scale employed in this study facilitates clear and quantifiable performance measurement due to its well-defined evaluation criteria, transparent weighting of indicators, and user-friendly application.

**Research Methodology**

The first step was to collect a dataset. For this research, the dataset with the relevant metrics to influence academic outcomes was collected from a data repository. Using Scikit Learn, the data was loaded and preprocessed in order to get a normalized dataset and the data features are modified to achieve the suitable format. Machine learning modules requires numeric inputs so it is important to arrange the data in a numerical form by converting all text values into numerical form using the LabelEncoder() function to do this in order to prevent any problems at later stages. Often, features are not given as continuous values but categorical. For example a lecturer could have features ["punctual", "not punctual"], ["regular", "not regular"], ["Physics",

"Computer Science"], ["yes", "no"]. Such features can be efficiently encoded as integers, for instance ["punctual", "regular", "Physics", "yes"] could be expressed as [0,1,3] while ["not punctual", "not regular", "Computer Science", "no"] would be [1,4,7].

The dataset as depicted in table 1 used in this study was obtained to predict instructor performance. It consists of 5000 rows and 12 columns. This study utilizes an evaluation system composed of twelve (12) performance metrics. From the performance metrics designed as seen in table 2 are all Likert-type, values are all taken from 1-5 where 1,2,3,4,5 represents 'Poor', 'Fair', 'Good', 'Very Good', and 'Excellent' respectively while the possible values for Q4 are 'Yes' and 'No'. The data is split into 70% as the training set used to train the model and 30% as the testing set used to evaluate the performance of the model. The Multilayer Perceptron (MLP) is a type of artificial neural network, a foundational model in machine learning, consisting of multiple layers of interconnected nodes (neurons) used to learn complex patterns in data, particularly for tasks like classification and regression. MLP can predict lecturer's performance by learning patterns and relationships between input features such as 'Research Publications', 'Punctuality', 'Administrative Duties' and their performance outcomes such as 'Excellent', 'Very Good', 'Fair'.

The MLP includes three layers; Input Layer has a neuron for each metric such as Research Publications and Administrative Duties, Hidden Layer has several layers with non-linear activation functions to capture complex patterns while the Output Layer predicts performance categories such as 'Excellent', 'Very Good', and 'Fair'. In this study, the Input Layer consists of twelve performance metrics such as 'Research Publications', 'Years of Experience', and 'Number of Courses Taught'. The Hidden Layer1 consists of 64 neurons with ReLU activation and Hidden Layer2 has 32 neurons with RELU activation while the Output Layer consists of 'Excellent', 'Very Good', 'Good', 'Fair', and 'Poor'. Then the model is trained using Adam optimizer and the parameters are fine-tuned in order the find the best set of parameters and achieve accuracy and minimal loss. Finally the model is tested using Mean Squared Error (MSE) and Mean Absolute Error (MAE) to assess how well the MLP performs on unseen data. Once trained and optimized, the model predicts the performance of lecturers based on real-time data. Figure 1 shows the flow of data in the system.

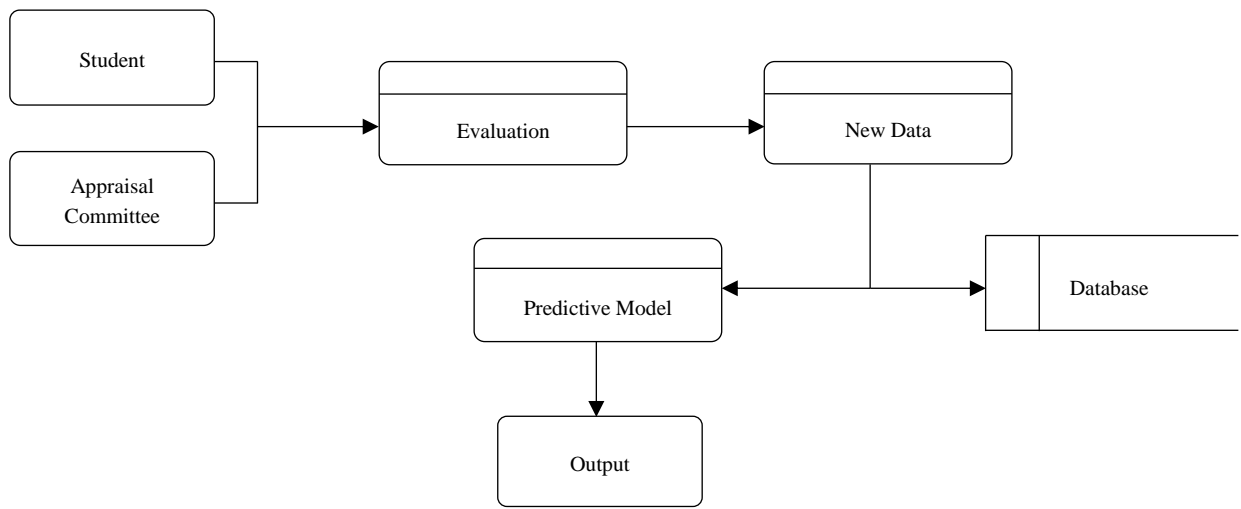

Figure 1: Dataflow of the System

Table 1 – Sample Dataset

| Lecturer ID | Student Evaluation Score | Research Publications | Teaching (Years) Experience | Administrative Duties |
|---|---|---|---|---|
| 101 | 85 | 5 | 10 | Yes |
| 102 | 75 | 2 | 8 | No |
| 103 | 90 | 8 | 12 | Yes |
| 104 | 65 | 1 | 5 | No |
| 105 | 78 | 4 | 7 | Yes |

Table 2 – Performance Metrics

| Appraisal Committee | Possible Values | Students | Possible Values |
|---|---|---|---|
| Research Publications | 1,2,3,4,5 | Punctuality | 1,2,3,4,5 |
| Years of Experience | 1,2,3,4,5 | Gave interesting and informative class | 1,2,3,4,5 |
| Number of Courses Taught | 1,2,3,4,5 | How regular is he/she to class | 1,2,3,4,5 |
| Administrative Duties | 1,2,3,4,5 | Was receptive to questions from students | 1,2,3,4,5 |
|  |  | Stimulated discussion on the subject | 1,2,3,4,5 |
|  |  | Stimulated me to think and learn | 1,2,3,4,5 |
|  |  | Was available to answer questions in office hours | 1,2,3,4,5 |
|  |  | Stimulated interest in the subject taught | 1,2,3,4,5 |

## Results

During the initial epochs, the model exhibited high loss values, with a starting MAE of approximately 74.92 and MSE of 5840.22. As the model progressed through additional epochs, it gradually minimized the errors, achieving a substantial reduction by the final epoch. By the end of training, the model's performance on the test set demonstrated an MAE of 13.76 and an MSE (Test Loss) of 256.99. These metrics are critical in accuracy on previously unseen data. Mean Absolute Error (MAE) metric at 13.76 on the test set, indicates that the model's average error in prediction, shows that the model can provide a fairly accurate estimation of scores. Figure 2 displays the Mean Absolute Error (MAE) for both training and validation data over 50 epochs. MAE is a measure of error between the predicted and actual values. A decrease in MAE over time indicates that the model is learning and improving its accuracy. In this case, both training and validation MAE decrease rapidly at the start and then stabilize, showing that the model has reached a point where additional training offers diminishing returns. Figure 3 shows the training and validation loss, calculated as Mean Squared Error (MSE), over 50 epochs. MSE gives more weight to larger errors, making it a good metric for regression problems. The MSE of 256.99 represent squared error. A Lower MSE value indicates higher accuracy and a value of 256.99 demonstrates that the model has effectively reduced prediction errors. The loss decreases significantly in the first few epochs and stabilizes as training continues, indicating that the model has learned to generalize from the data. The model also demonstrated an estimated accuracy rate of approximately 96%, validated its effectiveness in predicting lecturer performance.

## Conclusion

The development of a Performance Evaluation System marks a substantial advancement in the academic evaluation process for Nigerian lecturers. The platform has proven effective in automating and streamlining the evaluation procedures, providing a structured evaluations from Appraisal Committee and students.. By leveraging advanced algorithms and a user-friendly interface, the system ensures accurate, fair, and efficient evaluations, which enhances the overall quality of academic evaluations. The integration of a machine learning algorithm and real-time performance analytics improved the accuracy of evaluations and significantly reduced the time and manual effort traditionally required. Future work could with the aid of artificial intelligence system, analyze evaluation results and identify specific areas for improvement for each lecturer. After every evaluation process, it should provide constructive feedback to the evaluated lecturer highlighting areas requiring attention and offering suggestions for improvements. Additionally, it could integrate Natural Language Processing (NLP) to assess the emotional tone of feedback, such as identifying positive, negative, or neutral sentiments, providing a clearer understanding of student satisfaction or concerns. Additionally, NLP techniques could be employed to detect biased, repetitive, or irrelevant evaluations, ensuring the analysis emphasizes authentic and constructive input.